\documentclass[12pt]{article}
\usepackage{latexsym}
\usepackage{graphicx}
\usepackage{caption}
\usepackage{psfrag}
\usepackage{amsmath,amssymb,dsfont}
\newcommand{\be}{\begin{equation}}
\newcommand{\ee}{\end{equation}}
\newcommand{\bea}{\begin{eqnarray}}
\newcommand{\eea}{\end{eqnarray}}

\newcommand{\nn}{\nonumber}

\input epsf

\begin{document}

\begin{titlepage}

\begin{flushright}
\end{flushright}
\vspace*{1.5cm}
\begin{center}
{\Large \bf A Rational Approach to Resonance Saturation \\[.5 cm]
in large-$N_c$ QCD}\\[3.0cm]

{\bf Pere Masjuan } and {\bf Santiago Peris}\\[1cm]

Grup de F{\'\i}sica Te{\`o}rica and IFAE\\ Universitat Aut{\`o}noma de Barcelona, 08193 Barcelona, Spain.\\[0.5cm]

\end{center}

\vspace*{1.0cm}

\begin{abstract}

We point out that resonance saturation in QCD can be understood in the large-$N_c$ limit from the
mathematical theory of Pade Approximants to meromorphic functions. These approximants are rational
functions which encompass any saturation with a finite number of resonances as a particular
example, explaining several results which have appeared in the literature.  We review the main
properties of Pade Approximants with the help of a toy model for the $\langle VV-AA\rangle$
two-point correlator, paying particular attention to the relationship among the Chiral Expansion,
the Operator Product Expansion and the resonance spectrum. In passing, we also comment on an old
proposal made by Migdal in 1977 which has recently attracted much attention in the context of
AdS/QCD models. Finally, we apply the simplest Pade Approximant to the $\langle VV-AA\rangle$
correlator in the real case of QCD. The general conclusion is that a rational approximant may
reliably describe a Green's function in the Euclidean, but the same is not true in the Minkowski
regime due to the appearance of unphysical poles and/or residues.

\end{abstract}

\end{titlepage}

 \section{Introduction}

The strong Chiral Lagrangian is a systematic organization of the physics in powers of momenta and
quark masses, but requires knowledge of the low-energy constants (LEC) to make reliable
phenomenological predictions. As with any other effective field theory, these LECs play the role of
coupling constants and contain the information which comes from the integration of the heavy
degrees of freedom not explicitly present in the Chiral Lagrangian (e.g. meson resonances).

At $\mathcal{O}(p^4)$ there are 10 of these constants \cite{GL}. Although at this order there is
enough independent information to extract the values for these constants from experiment, this will
hardly ever be possible at the next order, $\mathcal{O}(p^6)$, because the number of constants
becomes more than a hundred \cite{p6}. In the electroweak sector the proliferation of constants
appears already at $\mathcal{O}(p^4)$ \cite{ew}. Although in principle these low-energy constants
may be computed on the lattice, in practice this has only been accomplished in a few cases for the
strong Chiral Lagrangian at $\mathcal{O}(p^4)$, and only recently \cite{lattice}.

The large $N_c$ expansion \cite{largeN} stands out as a very promising analytic approach capable of
dealing with the complexities of nonperturbative QCD while, at the same time, offering a relatively
simple and manageable description of the physics. For instance, mesons are $q\overline{q}$ states
with no width, the OZI rule is exact and there is even a proof of spontaneous chiral symmetry
breaking \cite{CW}. Furthermore, its interest has recently received a renewed boost indirectly
through the connection of some highly supersymmetric gauge theories to gravity \cite{Maldacena},
although the real relevance of this connection for QCD still remains to be seen. However, in spite
of all this, the fact that no solution to large-$N_c$ QCD has been found keeps posing a serious
limitation to doing phenomenology. For instance, in order to reproduce the parton model logarithm
which is present in QCD Green's functions in perturbation theory, an infinity of resonances is
necessary whose masses and decay constants are in principle unknown.

On the other hand, QCD Green's functions seem to be approximately saturated by just  a few
resonances; a property which has a long-standing phenomenological support going all the way back to
vector meson dominance ideas \cite{VMD}, although it has never been properly understood. In a
modern incarnation, this fact translated into the very successful observation \cite{swiss} that the
strong LECs at $\mathcal{O}(p^4)$ seem to be well saturated  by the lowest meson in the relevant
channel,\footnote{This is less clear in the scalar channel, however. See Ref. \cite{GP06}} after
certain constraints are imposed on some amplitudes at high-energy in order to match the expected
behavior in QCD \cite{swiss2, Moussallam}. It was then realized that all these successful results
could be encompassed at once as an approximation to large-$N_c$ QCD consisting in keeping only a
\emph{finite} (as opposed to the original infinite) set of resonances in Green's functions. This
approximation to large-$N_c$ QCD has been termed Minimal Hadronic Approximation (MHA) \cite{MHA}
because it implements the minimal constraints which are necessary to secure the leading non-trivial
behavior at large energy of certain Green's functions through the marriage of the old resonance
saturation and the large-$N_c$ approximation of QCD. In recent years, a large amount of work has
been dedicated to studying the consequences of these ideas \cite{theworks}.

However, the high-energy matching with a finite set of resonances, first suggested in
\cite{swiss2}, makes it clear that the treatment is not amenable to the methods of a conventional
effective field theory. An effective field theory is an approximation for energies smaller than a
heavy particle's mass and, therefore, cannot deal with momentum expansions at infinity as in the
case of the Operator product Expansion (OPE). In other words, the fact that the set of resonances
in each channel is really infinite precludes the naive expansion at large momentum because there is
always a mass in the spectrum which is even larger. The sum over an infinite set of resonances and
the expansion for large momentum are operations which do not commute \cite{GP02}. In those Green's
functions containing a contribution from the parton model logarithm, this is made self-evident
since a naive expansion at large momentum can only produce powers and not a logarithm, which is why
large-$N_c$ QCD requires an infinity of resonances in the first place.

The problem can be delayed one power of $\alpha_s$ if one requires the use of the resonance
Lagrangian \cite{swiss} to be limited only to Green's functions which are order parameters of
spontaneous chiral symmetry breaking. These order parameters vanish to all orders in $\alpha_s$ in
the chiral limit\footnote{E.g., the two-point correlator $\langle VV-AA\rangle$.}  and, therefore,
avoid the presence of the parton model logarithm which, otherwise, would preclude from the outset
any matching to a finite number of resonances. However, the concept of a Lagrangian whose validity
is restricted only to a certain class of Green's functions has never been totally clear; and even
if the resonance Lagrangian is restricted by definition to order parameters, the problem surfaces
again in the presence of logarithmic corrections  from nontrivial anomalous dimensions, which make
the exact matching at infinite momentum impossible.

In a slightly different context, a somewhat similar observation was also made in Ref.
\cite{Lipartia}. In this paper it was observed that it is impossible to satisfy the large momentum
fall-off expected in large-$N_c$ QCD for the form factors which can be defined through a
three-point Green's function, if the sum over resonances in the Green's function is restricted to a
finite set. Interestingly, this again pointed to an incompatibility of the QCD short-distance
behavior with an approximation to large $N_c$ which only kept a finite number of resonances.

A further piece of interesting evidence results from the comparison between the analysis in Refs.
 \cite{Friot} and \cite{swiss2}. After imposing some good high-energy behavior in several Green's
functions and form factors including, in particular, the axial form factor governing the decay
$\pi\rightarrow \gamma e\overline{\nu}$, Ref. \cite{swiss2} obtains, keeping only one vector state
$V$ and one axial-vector state $A$, that their two masses must be constrained by the relation
$M_A=\sqrt{2}M_V$. The work in Ref. \cite{Friot}, on the contrary, does not use the axial form
factor and obtains, after performing a very good fit within the same set of approximations, the
precise values $M_V=775.9\pm 0.5\ \mathrm{MeV}$ and $M_A=938.7\pm 1.4\ \mathrm{MeV}$. These values
for the masses, although close, are not entirely compatible with the previous relation. In other
words, the short-distance constraint from the axial form factor is not fully compatible with the
short-distance constraints used in \cite{Friot} if restricted to only one vector and one
axial-vector states\footnote{Adding one further state does not change the conclusion
\cite{Friot}.}.

In this paper we would like to suggest that all the above properties can be understood if the
approximation to large $N_c$ QCD with a finite number of resonances is reinterpreted within the
mathematical Theory of Pade Approximants (PA) to meromorphic functions \cite{Baker}. QCD Green's
functions in the large $N_c$ and chiral limits have an analytic structure in the complex momentum
plane which consists of an infinity of isolated poles but no cut, i.e. they become meromorphic
functions \cite{meromorphic}. As such, they have a well-defined series expansion in powers of
momentum around the origin with a finite radius of convergence given by the first resonance
mass\footnote{The pion pole can always be eliminated multiplying by enough powers of momentum. We
are assuming here the existence of a nonvanishing gap in large-$N_c$ QCD.}. This is all that is
needed to construct a Pade Approximant. A theorem by Pommerenke \cite{Pommerenke} assures then
convergence of any near diagonal PA to the true function for any finite momentum, over the whole
complex plane, except perhaps in a zero-area set. The poles of the original Green's function (i.e.
the resonance masses) belong to this zero-area set because not even the original function is
defined there, but there are also \emph{extra} poles. These extra poles are called ``defects'' in
the mathematical literature \cite{Baker}. When the Green's function being approximated is of the
Stieltjes type\footnote{Roughly this means that the associated spectral function is positive
definite, like in the case of the two-point correlator $\langle VV\rangle$. See Ref. \cite{Baker}
(chapter 5) for a more precise definition.}, the poles of the PA are always real and located on the
Minkowski region $\mathrm{Re}(q^2)=\mathrm{Re}(-Q^2)>0$, approaching the physical poles as the
order of the PA is increased \cite{PerisPade}. However, this takes place in a hierarchical way and,
while the poles in the PA which are closest to the origin are also very close to the physical
masses, the agreement quickly deteriorates and one may find that the last poles are several times
bigger than their physical counterparts \cite{PerisPade2}. The same is true of the residues. In
section 3, we will see with the help of a model that the same properties are met in a meromorphic
function whose spectral function is not positive definite, except that some of the poles in the PA
may even be complex.

This means that Minkowskian properties, such as masses and decay constants, cannot be reliably
determined from a PA except, perhaps, from the first poles which are closest to the origin. If not
all the residues and/or masses are physical, then there is no reason why they should be the same in
the form factor governing $\pi\rightarrow \gamma e\overline{\nu}$ and in the Green's function
$\langle VV-AA\rangle$, explaining the different results found in \cite{swiss2} and \cite{Friot} we
alluded to above. Furthermore, the form factors of all but the lightest mesons, defined through the
residues of the corresponding 3-point Green's functions, will not be reliably determined from a PA
to that Green's function, again in agreement with the findings in \cite{Lipartia}.

The situation in the Euclidean is different. In general, PAs cannot be expanded at infinite
momentum to generate an OPE type expansion for the true function. Nevertheless, Pommerenke's
theorem assures a good approximation at any finite momentum, no matter how large. Of course, the
order of the PA will have to increase, the larger the momentum region one wishes to approximate.
For instance, in Ref. \cite{PerisPade} it was shown with the help of a simple model how, even in
the case of the $\langle VV \rangle$ correlator which contains a $\log Q^2$ at large values of
$Q^2>0$, the PAs are capable of approximating the true function at any arbitrarily large (but
finite) value of $Q^2>0$, \emph{without} the need for a perturbative continuum. In section 3 we
will show, again with the help of a model, how this is also true in the more general case of a
non-positive definite spectral function such as $\langle VV-AA\rangle$. This means that PAs are a
reliable way to approximate the original Green's function in the Euclidean but not in the
Minkowskian regime.

In 1977, A.A. Migdal \cite{Migdal} suggested PAs as a method to extract the spectrum of large-$N_c$
QCD from the leading term in the OPE of the $\langle VV\rangle$ correlator, i.e. from the parton
model logarithm. However, nowadays this proposal should be considered unsatisfactory for a number
of reasons \cite{oscar}, the most simple of them being that different spectra may lead to the same
parton model logarithm \cite{spectra}. In fact, the full OPE series is expected to be only an
asymptotic expansion at $Q^2=\infty$ (i.e. with zero radius of convergence), and PAs constructed
from this type of expansions cannot in general reproduce the position of the physical poles
 \cite{Baker3}. For instance, we show this explicitly with the help of a model for $\langle
VV-AA\rangle$ in the Appendix. Migdal's approach has been recently adopted (in disguise) in some
models exploiting the so-called AdS/QCD correspondence \cite{Erlich} and, consequently, the same
criticism also applies to them.

In Ref. \cite{duality} a model for the $\langle VV-AA\rangle$ two-point correlator with a spectrum
consisting of an infinity of resonances was suggested as a theoretical laboratory for studying the
relationship between the spectrum and the coefficients of the OPE. In this paper several
conventional methods usually employed in the literature were tested against the exact result from
the model. These included: Finite Energy Sum Rules as in Ref. \cite{FESR}, pinched weights as in
Ref. \cite{pw}, Laplace transforms as in Ref. \cite{Narison} and, finally, also resonance
saturation as in the MHA method. The bottom line  was that no method was able to produce very
accurate predictions for the OPE coefficients. In all the methods but the last one, the reason for
this lack of accuracy  was basically due to the fact that the OPE requires an integral over the
whole spectrum, whereas  the integral is actually cut off at an upper limit (in the real case, the
upper limit is $m_{\tau}$). This is why even if one uses the real spectrum the result may be
inaccurate \cite{Donoghue}. In the case of the MHA the reason was, as we will comment upon below,
that the poles were not allowed to be complex.

In section 3 we will revisit this $\langle VV-AA\rangle$ model, now from the point of view of PAs.
The model reproduces the power behavior of QCD at large $Q^2>0$ except that the model is simple
enough not to have any $\log Q^2$ and, therefore, it cannot reproduce the nonvanishing anomalous
dimensions which exist in QCD. We do not think this is a major drawback, however, because in QCD
these logarithms are always screened by at least one power of $\alpha_s$ and, hence, in an
approximate sense, it may be licit to ignore them. In the model such an approximation becomes
exact\footnote{For a model with a $\log Q^2$, the reader may consult Ref. \cite{PerisPade}.}. Will
the PAs be able to reproduce the large $Q^2$ expansion of the $\langle VV-AA\rangle$ model? We will
see that the answer is affirmative. Therefore, the reason why the MHA method was not able to
predict accurately the OPE coefficients in Ref. \cite{duality} is because the lowest PA has complex
poles which were not allowed in \cite{duality}. When these complex poles are considered, the
accuracy achieved is better and, most importantly, improves for a higher PA. Since the model allows
the construction of PAs of a very high order, we have checked this convergence up to the Pade
$P^{50}_{52}$, which is able to reproduce the first non vanishing coefficient of the OPE in the
model with an accuracy of 52 decimal figures. Together with other numerical examples which will be
discussed in section 3, we take this as a clear evidence of the convergence of the method. This
renders some confidence that PAs may also do a good job in the real case of QCD.

The rest of the paper is organized as follows. In section 2 we review some generalities of rational
approximants, in section 3 we describe the $\langle VV-AA\rangle$ model and apply different
rational approximants to learn about the possible advantages and disadvantages of them. In section
4 we apply the simplest PA to the case of the real $\langle VV-AA\rangle $ two-point function in
QCD. Finally, we close with some conclusions.

 \section{Rational approximations: generalities}

Let a function $f(z)$ have an expansion around the origin of the complex plane of the form
\begin{eqnarray}\label{one}
    f(z)&=& \sum_{n=0}^{\infty} f_n z^n\quad , \quad z\rightarrow 0\ .
   \end{eqnarray}
One defines a Pade Approximant (PA) to $f(z)$ , denoted by $P^{M}_{N}(z)$, as a ratio of two
polynomials $Q_M(z), R_N(z)$\footnote{Without loss of generality we define, as it is usually done,
$R_N(0)=1$.}, of order $M$ and $N$ (respectively) in the variable z, with a \emph{contact} of order
$M+N$ with the expansion of $f(z)$ around $z=0$. This means that, when expanding $P^{M}_{N}(z)$
around $z=0$, one reproduces exactly the first $M+N$ coefficients of the expansion for $f(z)$ in
Eq. (\ref{one}):
\begin{equation}\label{two}
    P^{M}_{N}(z)=\frac{Q_M(z)}{R_N(z)}
    \approx f_0 + f_1\ z + f_2\ z^2 +...+ f_{M+N}\ z^{M+N}+ \mathcal{O}(z^{N+M+1})\ .
\end{equation}
At finite $z$, the rational function $P^{M}_{N}(z)$ constitutes a resummation of the series
(\ref{one}). Of special interest for us will be the case when $N=M+k$, for a fixed $k$, because
then the function behaves like $1/z^k$ at $z=\infty$. The corresponding PAs $P^{M}_{M+k}(z)$ belong
to what is called the near-diagonal sequence for $k\neq 0$, with the case $k=0$ being the diagonal
sequence.

The convergence properties of the PAs to a given function are much more difficult than those of
normal power series and this is an active field of research in Applied Mathematics. In particular,
those which concern meromorphic functions\footnote{A function is said to be meromorphic when its
singularities are only isolated poles.} are rather well-known and will be of particular interest
for this work. The main result which we will use is Pommerenke's Theorem \cite{Pommerenke} which
asserts that the sequence of (near) diagonal PA's to a meromorphic function is convergent
everywhere in any compact set of the complex plane except, perhaps, in a set of zero area. This set
obviously includes the set of poles where the original function $f(z)$ is clearly ill-defined but
there may be some other extraneous poles as well. For a given compact region in the complex plane,
the previous theorem of convergence requires that, either these extraneous poles move very far away
from the region as the order of the Pade increases, or they pair up with a close-by zero becoming
what is called a \emph{defect} in the mathematical jargon \cite{Baker2}. These are to be considered
artifacts of the approximation. Near the location of these extraneous poles the PA approximation
clearly breaks down but, away from these poles, the approximation is safe.

In the physical case the original function $f(z)$ will be a Green's function $G(Q^2)$ of the
momentum variable $Q^2$. In QCD in the large $N_c$ limit this Green's function is meromorphic with
all its poles located on the negative real axis in the complex $Q^2$ plane. These poles are
identified with the meson masses. On the other hand, the region to be approximated by the PAs will
be that of euclidean values for the momentum, i.e. $Q^2>0$. The expansion of $G(Q^2)$ for $Q^2$
large and positive coincides with the Operator Product Expansion.

In general a meromorphic function does not obey any positivity constraints and, as we will see,
this has as a consequence that some of the poles and residues of the PAs may become
\emph{complex}\footnote{A special case which does obey positivity constraints is when the function
is Stieltjes. In this case the poles and residues of the PAs are purely real and with the same sign
as those of the original function \cite{PerisPade}.}. This clearly precludes any possibility that
these poles and residues may have anything to do with the physical meson masses and decay
constants. However, and this is very important to realize, this does \emph{not} spoil the validity
of the rational approximation provided the poles, complex or not, are not in the region of $Q^2$
one is interested in. It is to be considered rather as the price to pay for using a rational
function, which has only a finite number of poles, as an approximation to a meromorphic function
with an infinite set of poles.

When the position of the poles in the original Green's function is known, at least for the lowest
lying states, it is interesting to devise a rational approximation which has this information
already built in. The corresponding approximants are called  Partial Pade Approximants (PPAs) in
the mathematical literature \cite{PGV} and are given by a rational function
$\mathbb{P}^{M}_{N,K}(Q^2)$:
\begin{equation}\label{three}
    \mathbb{P}^{M}_{N,K}(Q^2)=\frac{Q_M(Q^2)}{R_{N}(Q^2)\ T_{K}(Q^2)}\ ,
\end{equation}
where $Q_{M}(Q^2),R_{N}(Q^2)$ and $T_{K}(Q^2)$ are polynomials of order $M, N$ and $K$
(respectively) in the variable $Q^2$. The polynomial $T_{K}(Q^2)$ is defined by having $K$ zeros
precisely at the location of the lowest lying poles of the original Green's function\footnote{For
simplicity, we will assume that all the poles are simple.} i.e.
\begin{equation}\label{threeprime}
    T_{K}(Q^2)= (Q^2+ M_1^2)\ (Q^2+ M_2^2)\ ...\ (Q^2+ M_K^2) \ .
\end{equation}
As before the polynomial $R_{N}(Q^2)$ is chosen so that $R_{N}(0)=1$ and, together with $Q_M(Q^2)$,
they are defined so that the ratio $\mathbb{P}^{M}_{N,K}(Q^2)$ matches exactly the first $M+N$
terms in the expansion of the original function around $Q^2=0$, i.e. :
\begin{equation}\label{four}
    \mathbb{P}^{M}_{N,K}(Q^2)\approx f_0 + f_1\ Q^2 + f_2\ Q^4
    +...+ f_{M+N}\ Q^{2M+2N}+ \mathcal{O}(Q^{2N+2M+2})\ .
\end{equation}
At infinity, the PPA in Eq. (\ref{three}) obviously falls off like $1/Q^{2N+2K-2M}$. Exactly as it
happens in the case of PAs, also  the PPAs will have complex poles for a general meromorphic
function, which prevents it from any interpretation in terms of meson states.

Finally, another rational approximant defined in mathematics is the so-called Pade Type Approximant
(PTA) \cite{PGV} $\mathbb{T}^M_N(Q^2)$ :
\begin{equation}\label{five}
    \mathbb{T}^M_N(Q^2)=\frac{Q_M(Q^2)}{T_{N}(Q^2)}\ ,
\end{equation}
where $T_{N}(Q^2)$ is also given by the polynomial (\ref{threeprime}), now with $N$ preassigned
zeros at the corresponding position of the poles of the original Green's function, $G(Q^2)$. The
polynomial $Q_M(Q^2)$ is defined so that the expansion of the PTA around $Q^2=0$ agrees with that
of the original function up to and including terms of order $M+1$, i.e.
\begin{equation}\label{six}
    \mathbb{T}^M_N(Q^2)\approx f_0 + f_1\ Q^2 + f_2\ Q^4
    +...+ f_{M}\ Q^{2M}+ \mathcal{O}(Q^{2M+2})\ .
\end{equation}
At large values of $Q^2$, one has that $\mathbb{T}^M_N(Q^2)$ falls off like $1/Q^{2N-2M}$. Clearly
the PTAs are a particular case of the PPAs, i.e. $\mathbb{T}^M_N(Q^2)=\mathbb{P}^M_{0, N}(Q^2)$ and
coincide with what has been called the Hadronic Approximation to large-$N_c$ QCD in the literature
\cite{MHA}.

Let us summarize the mathematical jargon. A Pade Type Approximant (PTA) is a  rational function
with all the poles chosen in advance precisely at the physical masses. A Pade Approximant (PA) is
when all the poles are left free. The intermediate situation, with some poles fixed at the physical
masses and some left free, corresponds to what is called a Partial Pade Approximant (PPA).

\section{Testing rational approximations: a model}

Let us consider the two-point  functions of vector and axial-vector currents in the chiral limit
\be \label{correlator} \Pi^{V,A}_{\mu\nu}(q)=\ i\,\int d^4x\,e^{iqx}\langle J^{V,A}_{\mu}(x)
J^{\dag\ V,A}_{\nu}(0)\rangle = \left(q_{\mu} q_{\nu} - g_{\mu\nu} q^2 \right)\Pi_{V,A}(q^2) \ ,
\ee with $J_{V}^\mu(x) = {\overline d}(x)\gamma^\mu u(x)$ and $J_A^\mu(x) = {\overline
d}(x)\gamma^\mu \gamma^5 u(x)$. As it is known, the difference $\Pi_{V}(q^2)-\Pi_{A}(q^2)$
satisfies an unsubtracted dispersion relation\footnote{The upper cutoff which is needed to render
the dispersive integrals mathematically well defined can be sent to infinity provided it respects
chiral symmetry \cite{GP02}.} \be \label{dispersion} \Pi_{V-A}(q^2)= \int_0^{\infty}
\frac{dt}{t-q^2-i\epsilon}\ \frac{1}{\pi}\ {\rm Im}\,\Pi_{V-A}(t)\ . \ee

Following Refs. \cite{Shifman,duality}, we define our model by giving the spectrum as\bea
\label{spectrum} \frac{1}{\pi}\ {\rm Im}\,\Pi_V(t)&=& 2 F_{\rho}^2
\delta(t-M_{\rho}^2) + 2 \sum_{n=0}^{\infty} F^2_V(n)\delta(t-M^2_V(n))\ ,\nonumber\\
\frac{1}{\pi}\  {\rm Im}\,\Pi_A(t)&=& 2 F_{0}^2 \delta(t) + 2 \sum_{n=0}^{\infty}
F^2_A(n)\delta(t-M^2_A(n))\ . \eea Here $F_\rho,M_\rho $ are  the electromagnetic decay constant
and mass of the $\rho$ meson and $F_{V,A}(n)$ are the electromagnetic decay constants of the $n-th$
resonance in the vector (resp. axial) channels, while $M_{V,A}(n)$ are the corresponding masses.
$F_0$ is the pion decay constant in the chiral limit. The dependence on the resonance excitation
number $n$ is the following:
\begin{equation}\label{twoprime}
    F^2_{V,A}(n)=F^2= \mathrm{constant}\  ,
    \quad\quad M_{V,A}^2(n) = m_{V,A}^2 + n \ \Lambda^2\ ,
\end{equation}
in accord with known properties of the large-$N_c$ limit of QCD \cite{largeN} as well as alleged
properties of the associated Regge theory \cite{Regge}.

The combination
\begin{equation}\label{combo}
    \Pi_{LR}(q^2)=\frac{1}{2}(\Pi_V(q^2)-\Pi_A(q^2))
\end{equation}
thus reads
\begin{equation}\label{oneprime}
    \Pi_{LR}(q^2)=\frac{F^2_0}{q^2}+\frac{F_{\rho}^2}{-q^2+M_{\rho}^2}+
    \sum_{n=0}^{\infty} \left\{\frac{F^2}{-q^2+M^2_V(n)}-
    \frac{F^2}{-q^2+M^2_A(n)}\right\}\ .
\end{equation}
This two-point function can be expressed in terms of the Digamma function
$\psi(z)=\frac{d}{dz}\log{\Gamma(z)}$ as \cite{duality}
\begin{equation}\label{onecompact}
    \Pi_{LR}(q^2)=\frac{F^2_{0}}{q^2}+\frac{F_{\rho}^2}{-q^2+M_{\rho}^2}+
    \frac{F^2}{\Lambda^2}\left\{\psi\left(\frac{-q^2+m_A^2}{\Lambda^2}\right)-
    \psi\left(\frac{-q^2+m_V^2}{\Lambda^2}\right)\right\}\ .
\end{equation}
To resemble the case of QCD, we will demand that the usual parton-model logarithm is reproduced in
both vector and axial-vector channels and that the difference (\ref{dispersion}) has an operator
product expansion which starts at dimension six. A set of parameters satisfying these conditions is
given by\footnote{These numbers have been rounded off for the purpose of presentation. Some of the
exercises which will follow  require much more precision than the one shown here.}
\begin{eqnarray}\label{nature}
\hspace{-2. cm} F_{0}= 85.8 \,\,{\mathrm{MeV}}\ , \quad F_{\rho}= 133.884 \,\,{\mathrm{MeV}}\ ,
\quad F= 143.758 \,\,{\mathrm{MeV}}\, ,\qquad \qquad \\
M_{\rho}= 0.767 \,\,{\mathrm{GeV}}, \quad m_A= 1.182 \,\,{\mathrm{GeV}}, \quad m_V= 1.49371
\,\,{\mathrm{GeV}}\, , \quad \Lambda= 1.2774 \,\, {\mathrm{GeV}}\ , \nonumber
\end{eqnarray}
and is the one we will use in this section. This set of parameters has been chosen to resemble
those of the real world, while keeping the model at a manageable level. For instance, the values of
$F_{\rho}$ and $M_{\rho}$ in (\ref{nature}) are chosen so that the function $\Pi_{LR}$ in
(\ref{onecompact}) has vanishing $1/Q^2$ and $1/Q^4$ in the OPE at large $Q^2>0$, as in real QCD.
In fact, the model admits the introduction of finite widths (which is a $1/N_c$ effect) in the
manner described in Ref. \cite{Shifman}, after which the spectral function looks reasonably similar
to the experimental spectral function. This comparison can be found in Fig. 5 of Ref.
 \cite{duality}. But this model is also interesting for a very different reason. In Ref.
 \cite{duality} several attempts were made at determining the coefficients of the OPE by using the
methods which have become common practice in the literature. Among those we may list Finite Energy
Sum Rules \cite{FESR}, with pinched weights \cite{pw}, Laplace sum rules \cite{Narison} and Minimal
Hadronic Approximation \cite{MHA}. As it turned out, when these methods were tested on the model,
none of them was able to produce very accurate results. We think that this makes the model very
interesting (and challenging !) as a way to assess systematic errors \cite{almasy}.

Defining the expansion of the Green's function (\ref{dispersion}) in $Q^2=-q^2$ around
$Q^2=0,\infty$ as
\begin{equation}\label{exp}
    Q^2\ \Pi_{LR}(-Q^2)\approx \sum_{k} C_{2k}\ Q^{2k}\quad ,\quad \mathrm{with}\quad
    k=0, \pm 1, \pm 2, \pm 3, \ldots
\end{equation}
one obtains that the coefficients accompanying inverse powers of momentum, akin to the Operator
Product Expansion at large $Q^2>0$, are given by ($ p=1,2,3,...$ with $k=1-p$):
\begin{eqnarray}\label{ope}
     C_{2k} &=& -F_{0}^2 \ \delta_{p,1}+\nonumber \\
     &&\!\!\!\!\!\!\!\!\!\!\!\!
     (-1)^{p+1} \left[ F_{\rho}^2M_{\rho}^{2p-2}  -\frac{1}{p}F^2 \Lambda^{2p-2}
  \left\{B_{p}\left(\frac{m_V^2}{\Lambda^2}\right) -
  B_{p}\left(\frac{m_A^2}{\Lambda^2}\right)\right\}\right]\ ,
\end{eqnarray}
where $B_p(x)$ are the Bernoulli polynomials \cite{GP06}. As stated above, $F_{\rho}$ and
$M_{\rho}$ are defined by the condition that the above expression (\ref{ope}) vanishes for $k=0,-1$
enforcing that $ Q^2\ \Pi_{LR}(-Q^2)\sim Q^{-4}$ at large momentum, as in QCD. We emphasize that
the above coefficients of the OPE in Eq. (\ref{ope}) can not be calculated by a naive expansion at
large $Q^2$ of the Green's function in Eq. (\ref{oneprime}). In other words, physical masses and
decay constants do not satisfy the Weinberg sum rules \cite{GP02}.

On the other hand, for the coefficients accompanying nonnegative powers of momentum, akin to the
chiral expansion at small $Q^2$, one has ($k=1,2,3,...$):
\begin{equation}\label{L}
    C_{0}=-F_0^2\quad , \quad C_{2k}= (-1)^{k+1} \frac{F^2_{\rho}}{M^{2k}_{\rho}}- \frac{1}{(k-1)!}
    \frac{F^2}{\Lambda^{2k}}\left\{\psi^{(k-1)}\left(\frac{m_V^2}{\Lambda^2}\right)
    - \psi^{(k-1)}\left(\frac{m_A^2}{\Lambda^2}\right) \right\} \ ,
\end{equation}
where $\psi^{(k-1)}(z)= d^{k-1}\psi(z)/dz^{k-1}$. In Table \ref{table1} we collect the values for
the first few of these coefficients $C_{2k}$.

\begin{table}
\centering
\begin{tabular}{|c|c|c|c||c|c|c|}
  \hline
  $C_0$ & $C_2$ & $C_{4}$ & $C_6$ & $C_{-4}$ & $C_{-6}$ & $C_{-8}$ \\
  \hline
  $-7.362$ & $21.01$ & $-43.92$ & $81.81$ & $-2.592$ & $1.674$ & $-0.577$ \\
  \hline
\end{tabular}
\caption{\emph{Values of the coefficients $C_{2k}$ from the high- and low-$Q^2$ expansions of $Q^2\
\Pi_{LR}(-Q^2)$ in Eq. (\ref{exp}) in units of $10^{-3}\ GeV^{2-2k}$. Notice that $C_{-2}=0$ and
$C_0=-F_0^2$ (the pion decay constant in the chiral limit), see text.}}\label{table1}
\end{table}

Let us start with the construction of the rational approximants to the function $Q^2\
\Pi_{LR}(-Q^2)$. Since our original function (\ref{onecompact}) falls off at large $Q^2$ as
$Q^{-4}$, this is a constraint we will impose on all our approximants.

The simplest PA satisfying the right falloff at large momentum is $P^{0}_{2}(Q^2)$, so we will
begin with this case. In order to simplify the results, and unless explicitly stated otherwise, we
will assume that dimensionful quantities are expressed in units of $GeV$ to the appropriate power.
Fixing the three unknowns with the first three coefficients from the chiral expansion of
(\ref{onecompact}) (i.e. $C_{0,2,4}$) one gets the following rational function
\begin{equation}\label{P02}
    P^{0}_{2}(Q^2)=\frac{-\ r_R^2}{(Q^2+z_R) (Q^2+z_R^*)}\ , \quad r_R^2=3.379\times
    10^{-3}\ , \quad z_R=0.6550+i\ 0.1732\ .
\end{equation}
We can hardly overemphasize the striking appearance of a pair of complex-conjugate poles on the
Minkowski side of the complex $Q^2$ plane. Obviously, this means that these poles cannot be
interpreted in any way as the meson states appearing in the physical spectrum
(\ref{spectrum},\ref{oneprime}). In spite of this, if one expands (\ref{P02}) for large values of
$Q^2>0$, one finds $C_{-4}=-r_R^2=-3.379\times 10^{-3}$ which is not such a bad approximation for
this coefficient of the OPE, see Table \ref{table1}. Even better is the prediction of the fourth
term in the chiral expansion, which is $C_6=79.58\times 10^{-3}$.

This agreement is not a numerical coincidence and the approximation can be systematically improved
if more terms of the chiral expansion are known. In order to exemplify this, we have amused
ourselves by constructing the high-order PA $P^{50}_{52}(Q^2)$. This rational approximant correctly
determines the values for $C_{-4,-6,-8}$ with (respectively) 52, 48 and 45 decimal figures. In the
case of $C_{103}$, which is the first predictable term from the chiral expansion for this Pade, the
accuracy reaches some staggering 192 decimal figures. This is all in agreement with Pommerenke's
theorem \cite{Pommerenke}.

\begin{figure}
\renewcommand{\captionfont}{\small \it}
\renewcommand{\captionlabelfont}{\small \it}
\centering
\includegraphics[width=4.5in]{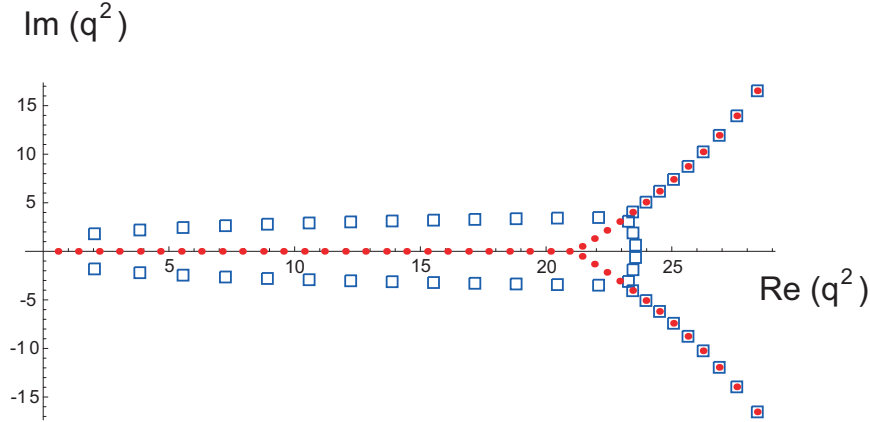}
\caption{Location of the poles (dots) and zeros (squares) of the Pade Approximant
$P^{50}_{52}(-q^2)$ in the complex $q^2$ plane.  We recall that $Q^2=-q^2$. Notice how zeros and
poles approximately coincide in the region which is farthest away from the origin. When the order
of the Pade is increased, the overall shape of the figure does not change but the two branches of
complex poles move towards the right, i.e. away from the origin.}\label{poles}
\end{figure}

As it happens for the PA (\ref{P02}), also higher-order PAs may develop some artificial poles. In
particular, Figure \ref{poles} shows the location of the 52 poles of the PA $P^{50}_{52}(Q^2)$ in
the complex $q^2$ plane. Of these, the first 25 are purely real and the rest are complex-conjugate
pairs. A detailed numerical analysis reveals that the poles and residues reproduce very well the
value of the meson masses and decay constants for the lowest part of the physical spectrum of the
model given in (\ref{oneprime}-\ref{nature}), but the agreement deteriorates very quickly as one
gets farther away from the origin, eventually becoming the complex numbers seen in Fig.
\ref{poles}. It is by creating these analytic defects that rational functions can effectively mimic
with a finite number of poles the infinite tower of poles present in the original function
(\ref{onecompact}).

For instance the values of the first pole and residue in $P^{50}_{52}(Q^2)$ reproduce those of the
$\rho$ in (\ref{nature}) within 193 astonishing decimal places for both. However, in the case of
the 25th pole, which is the last one still purely real, its location agrees with the physical mass
only with 3 decimal figures. This is not to be considered as a success, however, because after the
previous accuracy, this is quite a dramatic drop. In fact, the residue associated with this 25th
pole comes out to be 29 times the true value. The lesson we would like to draw from this exercise
should be clear: the determination of decay constants and masses extracted as the residues and
poles of a PA deteriorate very quickly as one moves away from the origin. There is no reason why
the last poles and residues in the PA are to be anywhere near their physical counterparts and their
identification with the particle's mass and decay constant should be considered unreliable.
Clearly, this particularly affects low-order PAs.

A very good accuracy can also be obtained in the determination of global euclidean observables such
as integrals of the Green's function over the interval $0\leq Q^2<\infty$. Notice that the region
where one approximates the true function is far away from the artificial poles in the PA. For
instance, one may consider the value for the integral
\begin{equation}\label{pi}
    I_{\pi}= \ (-1) \int_0^{\infty} dQ^2\ Q^2 \Pi_{LR}(Q^2)= 4.78719\times 10^{-3},
\end{equation}
which, up to a constant, would yield the electromagetic pion mass difference in the chiral limit
\cite{KPdeR} in the model (\ref{onecompact}). The PA $P^{50}_{52}(Q^2)$ reproduces the value for
this integral with more than 42 decimal figures. This suggests that one may use the integral
(\ref{pi}) as a further input to construct a PA.

For example if we fix the three unknowns in the PA $P^0_2(Q^2)$ by matching the first two terms
from the chiral expansion  but now we complete it with the pion mass difference (\ref{pi}) instead
of a third term from the chiral expansion as we did in (\ref{P02}),\footnote{We remark that this
procedure, although reasonable from the phenomenological point of view, strictly speaking lies
outside the standard mathematical theory of rational approximants \cite{Baker,PGV}.} the
approximant results to be
\begin{equation}\label{P02prime}
    \widetilde{P}^0_2(Q^2)=\frac{-\ r_R^2}{(Q^2+z_R) (Q^2+z_R^*)}\ ,\  \mathrm{with}
    \quad r_R^2=2.898\times 10^{-3}\ , \quad z_R=0.5618+i\ 0.2795\ .
\end{equation}
This determines $C_{-4}=-2.898\times 10^{-3}$ and $C_4=-41.26\times 10^{-3}$, which shows that
using the pion mass difference is not a bad idea. Notice how the position of the artificial pole
has changed with respect to (\ref{P02}).

Artificial poles and analytic defects are transient in nature, i.e. they appear and disappear from
a point in the complex plane when the order of the Pade is changed. On the contrary, the typical
sign that a pole in a Pade is associated with a truly physical pole is its stability under these
changes in the order of the Pade. Of course, when the order in the Pade increases there have to be
new poles by definition, and it is natural to expect that some of them will be defects. Pade
Approximants place some effective poles and residues in the complex $Q^2$ plane in order to mimic
the behavior of the true Green's function, but it can mimic the function only away from the poles,
e.g. in the Euclidean region. Obviously, PAs cannot converge at the poles,  in agreement with
Pommerenke's theorem \cite{Pommerenke}, since not even the true function is well defined there. The
point is  that what may look like a small correction in the Euclidean region may turn out to be a
large number in the Minknowski region. To exemplify this in simple terms, let us consider a very
small parameter $\epsilon$ and imagine that a given Pade $P(Q^2)$ produces the rational approximant
to the true Green's function $G(Q^2)$ given by
\begin{equation}\label{example}
    G(Q^2)\approx P(Q^2) \equiv R(Q^2) +\frac{\epsilon}{Q^2+M^2}\ ,
\end{equation}
where $R(Q^2)$ is the part of the Pade which is independent of $\epsilon$. Although for $Q^2>0$
there is a sense in which the last term is a small correction precisely because of the smallness of
$\epsilon$, for $Q^2<0$ this is no longer true because of the pole at $Q^2=-M^2$. This pole is in
general a defect and may not represent any physical mass. In fact, associated with this pole, there
is a very close-by zero of the Pade $P(Q^2)$ at $Q^2=-M^2-\epsilon\ {R(-M^2)}^{-1}$, as can be
immediately checked in (\ref{example}). This is another way of saying that a defect is
characterized by having an abnormally small residue and is the origin of the pairs of zeros and
poles in the y-shaped branches of Fig. \ref{poles}. Therefore, not only are defects unavoidable but
one could say they are even necessary for a Pade Approximant to approximate a meromorphic function
with an infinite set of poles.

Similarly to masses, also decay constants may be unreliable. To see this, imagine now that our Pade
is given by
\begin{equation}\label{example2}
    P(Q^2)= \frac{F}{Q^2+M^2}+ \frac{\epsilon}{(Q^2+M^2)\ (Q^2+M^2+\epsilon^2)}\
,
\end{equation}
again for a very small $\epsilon$. As before, the term proportional to $\epsilon$ may be considered
a small correction for $Q^2>0$. However, at the pole $Q^2=-M^2$ the decay constant becomes $F+
\epsilon^{-1}$ which, for $\epsilon$ small, may represent a huge correction. When the poles are
preassigned at the physical masses, like in the case of PTAs, it is the value of the residues that
compensates for the fact that the rational approximant lacks the infinite tower of resonances. As
we saw before, the residues of the poles in the Pade which lie farthest away from the origin are
the ones which get the largest distortion relative to their physical counterparts.

In real life, the number of available terms from the chiral expansion for the construction of a PA
is very limited.  Since the masses and decay constants of the first few vector and axial-vector
resonances are known, one may envisage the construction of a rational approximant having some of
its poles at the prescribed values given by the known masses of these resonances. If all the poles
in the approximant are prescribed this way (as in the MHA), we have a PTA. On the contrary, when
some of the poles are prescribed but some are also left free, then we have a PPA (see the previous
section).

Assuming that the first masses are known, let us proceed to constructing the PTAs (\ref{five}). The
lowest such PTA is $\mathbb{T}^{0}_{2}(Q^2)$, which contains two poles at the physical masses of
the $\rho$ and the first $A$ in the tower. Fixing the residue through the chiral expansion to be
$C_0=-F_0^2$, one obtains
\begin{equation}\label{PT02}
    \mathbb{T}^{0}_{2}(Q^2)=
\ \frac{-\ F_0^2 M^2_{\rho} M^2_{A} }{(Q^2+M^2_{\rho}) (Q^2+M^2_{A})}\ .
\end{equation}
Even though it has the same number of inputs ($C_0$ and the two masses), this rational approximant
does not do such a good job as the PAs (\ref{P02}) or (\ref{P02prime}). For instance, $C_{-4}$ is
2.3 times larger than the true value in Table \ref{table1}. As we have already stated, one way to
intuitively understand this result is the following. The OPE is an expansion at $Q^2= \infty$ and
therefore knows about the whole spectrum because no resonance is heavy enough with respect to $Q^2$
to become negligible in the expansion, i.e., the infinite tower of resonances does not decouple in
the OPE. Chopping an infinite set of poles down to a finite set may be a good approximation, but
only at the expense of some changes. These changes amount to the appearance of poles and residues
in the PA which the original function does not have. This is how the PA (\ref{P02}) manages to
approximate the true function (\ref{onecompact}). However, by construction, the PTA (\ref{PT02})
does not allow the presence of any artificial pole because, unlike in a PA, all its poles are fixed
at the physical values. Consequently, it only has its residues as a means to compensate for the
infinite tower of poles present in the true function and, hence, does a poorer job than the PA
(\ref{P02}), particularly in determining large-$Q^2$ observables like $C_{-4}$. Indeed, the role
played by the residues in the approximation can be appreciated by comparing the true values of the
decay constants to those extracted from (\ref{PT02}). Although the one of the $\rho$ is within 30\%
of the true value, that of the $A$ is off by 100\%.

A different matter is the prediction of low-energy observables such as, e.g., the chiral
coefficients. In this case heavy resonances make a small contribution and this means that the
infinite tower of resonances does decouple.\footnote{This is because the residues $F^2$ in the
Green's function (\ref{onecompact}) stay constant as the masses grow. This behavior does not hold
in the case of the scalar and pseudoscalar two-point functions \cite{GP06}.} Truncating the
infinite tower down to a finite set of poles is not such a severe simplification in this case,
which helps understand why a PTA may do a good job predicting unknown chiral coefficients. Indeed,
(\ref{PT02}) reproduces the value of $C_2$ within an accuracy of 15\%, growing to  22\% in the case
of $C_4$. A global observable like $I_{\pi}$ averages the low and the high $Q^2$ behaviors and ends
up differing from the true value (\ref{pi}) by 35\%. This gives some confidence that observables
which are integrals over Euclidean momentum may be reasonably estimated with MHA as, e.g.,  in the
$B_K$ calculation of Ref. \cite{BK}.

Improving on the PTA (\ref{PT02}) by adding in the first resonance mass from the vector tower
produces the following approximant
\begin{equation}\label{PT13}
    \mathbb{T}^1_3(Q^2)=\frac{a+b\ Q^2}{(Q^2+M^2_{\rho}) (Q^2+M^2_{A})
(Q^2+M^2_{V})}\ ,\  \mathrm{with}\ \left\{
                     \begin{array}{ll}
                       a =& -13.5\times 10^{-3},\cr
   b=& +1.33\times 10^{-4} \quad ,  \hbox{}
                     \end{array}
                   \right.
\end{equation}
where the values of the chiral coefficients $C_0$ and $C_2$ have been used to determine the
parameters $a$ and $ b$. The prediction for $C_4$ is much better now (only 2\% off), in agreement
with our previous comments. The prediction for $C_{-4}$ is still very bad, becoming now 19 times
smaller than the exact value. Nevertheless,  it eventually gets much better if PTAs of very high
order are constructed. For instance, we have found $C_{-4}=-2.58\times 10^{-3}$ for the approximant
$\mathbb{T}^{7}_{9}$ with 9 poles. Similarly, we have also checked that the prediction of the
chiral coefficients and the integral (\ref{pi}) improve with higher-order PTAs.

However, another matter is the prediction of the residues. For instance, the prediction for the
decay constant of the state with mass $M_V$ in (\ref{PT13}) is smaller than the exact value in the
model (\ref{nature}) by a factor of $2$. In general, we have seen that the residues of the poles
always deteriorate very quickly so that the residue corresponding to the pole which is at the
greatest distance from the origin is nowhere near the exact value. We again explicitly checked this
up to the approximant $\mathbb{T}^{7}_{9}$, in which case the decay constant for this pole is
almost 5 times smaller than the exact value. The conclusion, therefore, is that PTAs are able to
approximate the exact function only at the expense of changing the residues of the poles from their
physical values. Identifying residues with physical decay constants may be completely wrong in a
PTA for the poles which are farthest away from the origin.

As an intermediate approach between PAs and PTAs, there are the PPAs (\ref{three}) where some poles
are fixed at their physical values while some others are left free. The simplest of such rational
approximants is $\mathbb{P}^0_{1,1}(Q^2)$ (see the previous section for notation). Fixing its 3
unknowns with $M^2_{\rho}, C_0$ and $C_2$, one obtains
\begin{equation}\label{PP02}
    \mathbb{P}^0_{1,1}(Q^2)=\frac{-\ r_R^2}{(Q^2+M_{\rho}^2) (Q^2+z_R)}\ ,\ \mathrm{with} \quad
r_R^2= 3.75\times 10^{-3}   \ , \quad z_R=0.8665   \ .
\end{equation}
As can be seen, the mass (squared)  of the first $A$ resonance is predicted to be at $z_R$ which is
sensibly smaller than the true value in (\ref{nature})\footnote{Intriguinly enough, this is also
what happens in the real case of QCD \cite{swiss, Friot}.}. The rational function (\ref{PP02})
predicts $C_{-4}=-r_R^2=-3.75\times 10^{-3}$ which is a better determination than that of the PTA
(\ref{PT02}) with the same number of inputs, and $C_{4}=-45.52\times 10^{-3}$ which is not bad
either. Concerning the pion mass difference, one gets $I_{\pi}=5.22\times 10^{-3}$. However, as
compared to the PAs (\ref{P02}) or (\ref{P02prime}), the PPA (\ref{PP02}) does not represent a
clear improvement.

In order to improve on accuracy of the PPA, one may try to use the mass and decay constant of the
first resonance, $M_{\rho}$ and $F_{\rho}$, in addition to the pion mass difference and the chiral
coefficients $C_0, C_2$ and build the $\mathbb{P}_{2,1}^1(Q^2)$, which can be written as:
\begin{equation}\label{PPA}
    \mathbb{P}_{2,1}^1(Q^2)=\frac{F_{\rho}^2 M_{\rho}^2}{Q^2+ M_{\rho}^2}+
\frac{a- F_{\rho}^2 M_{\rho}^2\ Q^2}{(Q^2+z_c)\
    (Q^2+z_c^*)}\ ,\ \left\{
                     \begin{array}{ll}
                       a =& 17.43\times 10^{-3},\cr
   z_c=& 1.24+i\ 0.34 \quad .  \hbox{}
                     \end{array}
                   \right.
    \end{equation}
This PPA, upon expansion at large and small $Q^2$, determines $C_{-4}=- 2.47\times 10^{-3} $ and
$C_4= -44.0\times 10^{-3}$ to be compared with the corresponding coefficient in Table
{\ref{table1}. The accuracy obtained is better than that of (\ref{P02prime}), but this is probably
to be expected since (\ref{PPA}) has more inputs.

Based on the previous  numerical experiments done on the model in Eq.
(\ref{onecompact},\ref{nature}) (and many others), we now summarize the following conclusions.
Although, in principle, the PAs have the advantage of reaching the best precision by carefully
adjusting the polynomial in the denominator to have some effective poles which simulate the
infinite tower present in (\ref{onecompact}), they have the disadvantage that some of the terms in
the low-$Q^2$ expansion are required precisely to construct this denominator. This hampers the
construction of high-order PAs and consequently limits the possible accuracy.

When the locations of the first poles in the true function are known, there is the possibility to
construct PTAs (with all the poles fixed at the true values) and PPA (with some of the poles fixed
and some left free). As we have seen, although the PTA may approximate low-$Q^2$ properties of the
true function reasonably well, the large-$Q^2$ properties tend to be much worse, at least as long
as they are not of unrealistically high order. The PPAs, on the other hand, interpolate smoothly
between the PAs (only free poles) and the PTAs (no free pole). Depending on the case, one may
choose one or several of these rational approximants. However, common to all the rational
approximants constructed is the fact that the residues and/or poles which are farthest away from
the origin are in general unrelated to their physical counterparts.

\section{The QCD case}

Let us now discuss the real case of large-$N_c$ QCD in the chiral limit. In contrast to the case of
the previous model, any analysis in this case is limited by two obvious facts. First, any input
value will have an error (from experiment and because of the chiral and large-$N_c$ limits), and
this error will propagate through the rational approximant. And second, it is not possible to go to
high orders in the construction of rational approximants due to the rather sparse set of input
data. In spite of these difficulties one may feel encouraged by the phenomenological fact that
resonance saturation approximates meson physics rather well.

The simplest PA to the function $Q^2 \Pi_{LR}(-Q^2)$ with the right fall-off as $Q^{-4}$ at large
$Q^2$ is  $P^{0}_{2}(Q^2)$:
\begin{equation}\label{onepade}
    P^{0}_{2}(Q^2)=\frac{a}{1+ A \ Q^2 + B\ Q^4}\ .
\end{equation}
The values of the three unknowns $a,A$ and $B$ may be fixed by requiring that this PA reproduces
the correct values for $F_0, L_{10}$ and $I_{\pi}$ \footnote{Recall that $I_{\pi}$ is, up to a
constant, the electromagnetic pion mass difference $\delta m_{\pi}$ \cite{Friot} and is defined in
terms of $\Pi_{LR}$ as in Eq. (\ref{pi}).} given by
\begin{eqnarray}\label{realnature}
F_0&=& 0.086\pm  0.001\ \mathrm{GeV} \quad , \nn\\
   \delta m_{\pi}= 4.5936\pm 0.0005\ \mathrm{MeV}\
&\Longrightarrow  &\  I_{\pi}= (5.480 \pm 0.006)\times 10^{-3} \mathrm{GeV}^4\ , \nn \\
L_{10}(0.5\ \mathrm{GeV}) \leq L_{10}\leq L_{10}(1.1\ \mathrm{GeV})& \Longrightarrow &
L_{10}=\left( -5.13\pm 0.6\right) \times 10^{-3} \ .
\end{eqnarray}
The low-energy constant $L_{10}$ is related to the chiral coefficient $C_2$, in the notation of Eq.
(\ref{exp}), by $C_2=-4L_{10}$. Since $L_{10}$ does not run in the large-$N_c$ limit, it is not
clear at what scale to evaluate $L_{10}(\mu)$ \cite{subleading}. In Eq. (\ref{realnature}) we have
varied $\mu$ in the range $0.5\ \mathrm{GeV}\leq \mu \leq 1.1\ \mathrm{GeV}$ as a way to estimate
$1/N_c$ systematic effects. The central value corresponds to the result for $L_{10}(M_{\rho})$
found in Ref. \cite{L10}. The other results in (\ref{realnature}) are  extracted from Refs.
 \cite{GL,PDG}.

Obviously, the PA  (\ref{onepade}) can also be rewritten as
\begin{equation}\label{P02real}
     P^{0}_{2}(Q^2)=\frac{-\ r^2}{(Q^2+z_V) (Q^2+z_A)}\ ,
\end{equation}
in terms of two poles $z_{V,A}$. In order to discuss the nature of these poles, we will define the
dimensionless parameter $\zeta$ by the combination
\begin{equation}\label{zeta}
    \zeta\equiv  -4 L_{10}\ \frac{\ I_{\pi}}{F_0^4}= 2.06\pm 0.25\quad ,
\end{equation}
where the values in (\ref{realnature}) above have been used in the last step.
 Imposing the constraints (\ref{realnature}) on
the PA (\ref{P02real}) one finds two types of solutions depending on the value of
 $\zeta$: for $\zeta> 2$ the two poles $z_{V,A}$ are real, whereas for
$\zeta< 2$ the two poles are complex. At $\zeta=2$, the two solutions coincide. To see this, let us
write the set of equations satisfied by the PA (\ref{P02real}) as:
\begin{eqnarray}\label{set}
  F_0^2 &=& \frac{r^2}{z_V z_A} \nn \\
  -4L_{10} &=& F_0^2\left(\frac{1}{z_V}+\frac{1}{z_A} \right) \nn \\
  I_{\pi} &=& F_0^2\frac{z_V z_A}{z_A-z_V}\log\frac{z_A}{z_V}\ .
\end{eqnarray}
The first of these equations can be used to determine the value of the residue $r^2$ in terms of
$z_Vz_A$. In order to analyze the other two, let us first assume that both poles $z_{V,A}$ are
real. In this case, they also have to be positive  or else the integral $I_{\pi}$ will not exist
because it runs over all positive values of $Q^2$. Let us now make the change of variables
\begin{equation}\label{change}
    z_V=R\ (1- x) \qquad , \qquad  z_A=R\ (1+x)\quad .
\end{equation}
The condition $z_{V,A}>0$ translates into $R>0, |x|< 1$. In terms of these new variables, the
second and third equations in (\ref{set}) can be combined into
\begin{equation}\label{master}
    \zeta= \frac{1}{x}\log \frac{1+x}{1-x}\ ,
\end{equation}
where the definition (\ref{zeta}) for $\zeta$ has been used. With the help of the identity $\log
(1+x/1-x)=2\ \mathrm{th}^{-1}x$ (for $|x|< 1$), one can finally rewrite this expression as
\begin{equation}\label{master2}
    \zeta=\frac{2}{x}\ \mathrm{th}^{-1}x\quad , \quad (x\ \mathrm{real})
\end{equation}
which is an equation with a solution for $x$ only if $\zeta \geq 2$. Once this value of $x$ is
found, the value of $R$ can always be obtained from one of the last two equations (\ref{set}) and
this determines the two real poles $z_{V,A}$ from (\ref{change}).

On the other hand, when $\zeta <2$, Eq. (\ref{master2}) does not have a solution. However,
according to (\ref{zeta}), $\zeta$ can \emph{also} be smaller than 2. In order to study this case,
we may use the identity $\mathrm{th}^{-1}(i\ y)= i\ \tan^{-1} (y)$ to rewrite the above equation
(\ref{master2}) in terms of the variable $x=i\ y$ ($y$ real) as
\begin{equation}\label{master3}
    \zeta=\frac{2}{y}\ \mathrm{tan}^{-1}y\quad , \quad (y\ \mathrm{real}) .
\end{equation}
One now finds that this equation has a solution for $y$ when $\zeta \leq 2$. In this case the poles
of the PA (\ref{onepade}) are complex-conjugate to each other and can be obtained as
$z_{V,A}=R(1\pm i\ y)$. These poles, obviously, cannot be associated with any resonance mass and
this is why this solution has been discarded in all resonance saturation schemes up to now.
However, from the point of view of the rational approximant (\ref{onepade}) there is nothing wrong
with this complex solution, as the approximant is real and well behaved. From the lessons learned
in the previous section with the model, there is no reason to discard this solution since, as we
saw, rational approximants may use complex poles to produce accurate approximations. Therefore, we
propose to use \emph{both} the complex as well as the real solution for the poles $z_{V,A}$, at
least insofar as the value for $\zeta\gtrless 2$. In this case we obtain, using the values given in
Eqs. (\ref{realnature}),
\begin{eqnarray}
 &&\!\!\!\!\!\!\!\!\!\!\!\!\!\!\!\!\!\!\!\!\!\!\!\!\!(\zeta\geq 2)\ ,\quad   r^2 = -(4.1\pm 0.5)\times 10^{-3}\ ,\
   z_V = (0.77)^2\pm 0.15\  ,\ z_A = (0.96)^2 \pm 0.41 \label{results1}\\
  &&\!\!\!\!\!\!\!\!\!\!\!\!\!\!\!\!\!\!\!\!\!\!\!\!\!(\zeta \leq 2)\  ,\quad   r^2=-(3.9 \pm 0.1) \times 10^{-3}\ , \
    z_V=z_A^* = (0.66\pm 0.06)+ i\ (0.25\pm 0.25)\ \label{results2},
\end{eqnarray}
in units of $\mathrm{GeV}^6$ for $r^2$, and $\mathrm{GeV}^2$ for $z_{V,A}$. The two solutions in
Eqs. (\ref{results1},\ref{results2}) have been separated for illustrative purposes only. It is
clear that they are continuously connected through the boundary at $\zeta=2$, at which value the
two poles coincide and $z_V=z_A \simeq 0.72$. The errors quoted are the result of scanning the
spread of values  in (\ref{realnature}) through the equations (\ref{set}).

With both set of values in (\ref{results1},\ref{results2}), one can get to a prediction for the
chiral and OPE coefficients by expansion in $Q^2$ and $1/Q^2$, respectively. These expansions of
the PA can be done entirely in the Euclidean region $Q^2>0$, away from the position of the poles
$z_{V,A}$, whether real or complex. Recalling the notation in Eq. (\ref{exp}), the above
$P^{0}_{2}(Q^2)$ produces the coefficients for these expansions collected in Table \ref{table2}.
The values for the OPE coefficients $C_{-4,-6,-8}$ in this table are compatible with those of Ref.
 \cite{Friot}, after multiplying by a factor of two in order to agree with the normalization used by
these authors. However, the spectrum  in our case is different because of the complex solution in
(\ref{results2}). As we saw in the previous section with a model, this again shows that Euclidean
properties of a given Green's function, such as the OPE and chiral expansions, or integrals over
$Q^2>0$ are safer to approximate with a rational approximant than Minkowskian quantities, such as
resonance masses and decay constants.

\begin{table}
\centering
\begin{tabular}{|c|c|c|c|c||c|c|c|c|}
  \hline
  $C_0$ & $C_2$ & $C_{4}$ & $C_6$ & $C_8$ & $C_{-4}$ & $C_{-6}$ & $C_{-8}$  \\
  \hline
  $-F_0^2$ & $-4\,L_{10}$ & $-43\pm 13$ & $81\pm 53$ & $-145\pm 120$ & $-4.1\pm 0.5$
& $6\pm 2 $& $-7\pm 6$   \\
  \hline
\end{tabular}
\caption{\emph{Values of the coefficients $C_{2k}$ in the high- and low-$Q^2$ expansions of $Q^2\
\Pi_{LR}(-Q^2)$ in Eq. (\ref{exp}) in units of $10^{-3}\ GeV^{2-2k}$. Recall that $C_{-2}=0$.
}}\label{table2}
\end{table}

\section{Conclusions}

In this article we pointed out that approximating large-$N_c$ QCD with a finite number of
resonances may be reinterpreted within the mathematical Theory of Pade Approximants to meromorphic
functions \cite{Baker}.

The main results of this theory may be summarized as follows. One may expect convergence of a
sequence of Pade Approximants to a QCD Green's function in the large-$N_c$ limit in any compact
region of the complex $Q^2$  plane except at most in a zero-area set \cite{Pommerenke}. This set
without convergence comprises the poles of the original Green's function together with some other
artificial poles generated by the approximant which the original function does not have. As the
order of the PA grows, the previous convergence property implies that any given artificial pole
either goes to infinity, away from the relevant region, or is almost compensated by a nearby zero.
This symbiosis between a pole and a zero is called a defect. Although close to a pole the rational
approximation breaks down, in a region which is far away from it the approximation should work
well.

We have reviewed the main results of this theory with the help of a model for the two-point Green's
function $\langle VV-AA\rangle$. The simpler case of a Green's function of the Stieltjes type, such
as the two-point correlator $\langle VV\rangle$, was previously considered in Ref.
 \cite{PerisPade}. We have seen in the case of this particular  model how rational approximants
create the expected artificial poles (and the corresponding residues) in the Minkowski region
$\mathrm{Re}(q^2)>0$ while, at the same time, yielding an accurate description of the Green's
function in the Euclidean region $\mathrm{Re}(q^2)<0$. This happens in a hierarchical way: although
the first poles/residues  in a PA may be used to describe the physical masses/decay constants
reasonably well, the last ones give only a very poor description. Therefore, it is in general
unreliable to extract properties of individual mesons, such as masses and decay constants, from an
approximation to large-$N_c$ QCD with only a finite number of states. Since a form factor, like a
decay constant, is obtained as the residue of a Green's function at the corresponding pole(s), this
also means that one may not extract a meson form factor from a rational approximant to a 3-point
Green's function, in agreement with \cite{Lipartia}.  This observation may explain why the analysis
of Ref. \cite{ecker}, which is based on an extraction of matrix elements such as $\langle
\pi|S|P\rangle$ and $\langle \pi|P|S\rangle$ from the 3-point function $\langle SPP\rangle$, finds
values for the $K_{\ell 3}$ form factor which are different from those obtained in other
 analyses \cite{LRoos}.

In spite of all the above problems related to the Minkowski region, our model shows how Pade
Approximants may nevertheless be a useful tool in other regions of momentum space. We think that
this is also true in the real case of QCD in the large-$N_c$ limit. In this case one may use the
first few terms of the chiral and operator product expansions of a given Green's function to
construct a Pade Approximant which should yield a reasonable description of this function in those
regions of momentum space which are free of poles. In this construction, Pade Approximants
containing complex poles, if they appear, should not be dismissed.

We have also reanalyzed the simplest approximation to the $\langle VV-AA\rangle$ Green's function
in real QCD which consists of keeping only two poles, and we have found that, depending on the
value of the combination $\zeta$ in Eq. (\ref{zeta}), these two poles may actually be complex.

However, if not all the residues and masses in a rational approximant are physical, this poses a
challenge to any attempt to use a Lagrangian with a finite number of resonances such as, for
example, the ones in Ref. \cite{swiss,swiss2}, for describing Green's functions in the large-$N_c$
limit of QCD. Even if these Lagrangians are interpreted in terms of PTAs, with the poles fixed at
the physical value of the meson masses, we have seen how the residues then get very large
corrections with respect to their physical counterparts. Can these residues be efficiently
incorporated in a Lagrangian framework? We hope to be able to devote some work to answering this
and related questions in the future.

\vspace{1cm}

\textbf{Acknowledgements}

S.P. is indebted to M. Golterman, M. Knecht and E. de Rafael for innumerable discussions during the
last years which have become crucial to shape his understanding on these issues. He is also very
grateful to C. Diaz-Mendoza, P. Gonzalez-Vera and R. Orive, from the Dept. of Mathematical Analysis
at La Laguna Univ., for invaluable conversations on the properties of Pade Approximants as well as
for hospitality. We thank S. Friot, M. Golterman, M. Jamin, R. Kaiser, J. Portoles and E. de Rafael
for comments on the manuscript.

This work has been supported by CICYT-FEDER-FPA2005-02211, SGR2005-00916 and by the EU Contract No.
MRTN-CT-2006-035482, ``FLAVIAnet''.

\vspace{2cm}

 \textbf{\Large APPENDIX }

\vspace{1cm}

Here we will show how the PAs constructed from the OPE do not in general reproduce even the first
resonances in the spectrum, unlike those constructed from the chiral expansion. Again, we will use
the model of section 3 as an example. Recalling the definition of the OPE given in Eq. (\ref{exp}),
with the corresponding coefficients (\ref{ope}), it is straightforward to construct a PA in $1/Q^2$
around infinity, i.e. by matching powers of the OPE in $1/Q^2$. The construction parallels that in
Eq. (\ref{two}) but with the replacement $z= 1/Q^2$. Since the function $Q^2 \Pi_{LR} (-Q^2)$
behaves like a constant for $Q^2\rightarrow 0$, we will consider diagonal Pade Approximants, i.e.
of the form $P^{n}_{n}(1/Q^2)$, in order to reproduce this behavior. Figure \ref{poles2} shows the
position of the poles and zeros of the PA $P^{50}_{50}(-1/q^2)$ in the complex $q^2$ plane. As it
is clear from this plot, the positions of the poles have nothing to do with the physical masses in
the model, given by Eqs. (\ref{twoprime}-\ref{nature}), even for the lightest states. This is to be
contrasted with what happens with the PA constructed from the chiral expansion around $Q^2$, which
is shown in Fig. \ref{poles}. The difference between the two behaviors is due to the fact that,
while the chiral expansion has a finite radius of convergence, the radius of convergence of the OPE
vanishes because this expansion is asymptotic.

\begin{figure}
\renewcommand{\captionfont}{\small \it}
\renewcommand{\captionlabelfont}{\small \it}
\centering
\includegraphics[width=5in]{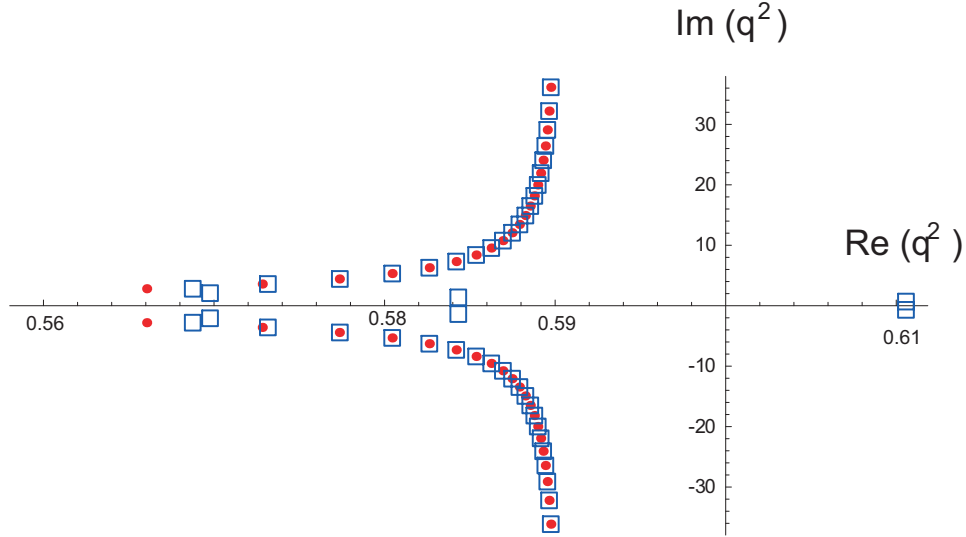}
\caption{Location of the poles (dots) and zeros (squares) of the Pade Approximant
$P^{50}_{50}(-1/q^2)$, constructed from the OPE of $Q^2 \Pi_{LR}$ in (\ref{onecompact}), in the
complex $q^2$ plane. We recall that $Q^2=-q^2$. The poles are all complex-conjugate
pairs.}\label{poles2}
\end{figure}


\begin{thebibliography}{99}


\bibitem{GL}
 J.~Gasser and H.~Leutwyler,
  Nucl.\ Phys.\ B {\bf 250}, 465 (1985).


\bibitem{p6}
 H.~W.~Fearing and S.~Scherer,
  Phys.\ Rev.\  D {\bf 53} (1996) 315
  [arXiv:hep-ph/9408346];
 J.~Bijnens, G.~Colangelo and G.~Ecker,
  JHEP {\bf 9902} (1999) 020
  [arXiv:hep-ph/9902437];
J.~Bijnens, L.~Girlanda and P.~Talavera,
  Eur.\ Phys.\ J.\  C {\bf 23} (2002) 539
  [arXiv:hep-ph/0110400].

\bibitem{ew}
J.~Kambor, J.~Missimer and D.~Wyler,
  Nucl.\ Phys.\  B {\bf 346} (1990) 17;
G.~Esposito-Farese,
  Z.\ Phys.\  C {\bf 50} (1991) 255.

\bibitem{lattice}
 C.~Bernard {\it et al.},
  arXiv:hep-lat/0611024 and references therein.


\bibitem{largeN}G.~'t Hooft,
Nucl.\ Phys.\ B {\bf 72} (1974) 461;
E.~Witten,
Nucl.\ Phys.\ B {\bf 160} (1979) 57.

\bibitem{CW}
 S.~R.~Coleman and E.~Witten,
  Phys.\ Rev.\ Lett.\  {\bf 45} (1980) 100.


\bibitem{Maldacena}
 J.~M.~Maldacena,
  Adv.\ Theor.\ Math.\ Phys.\  {\bf 2} (1998) 231
  [Int.\ J.\ Theor.\ Phys.\  {\bf 38} (1999) 1113]
  [arXiv:hep-th/9711200].



\bibitem{VMD}
J.J. Sakurai, ``\textit{Currents and Mesons}'', Chicago Lectures in Physics, The Univ. of Chicago
Press, 1969; see also  A.~Bramon, E.~Etim and M.~Greco,
  Phys.\ Lett.\  B {\bf 41} (1972) 609.


\bibitem{swiss}
J.~F.~Donoghue, C.~Ramirez and G.~Valencia,
  Phys.\ Rev.\  D {\bf 39} (1989) 1947;
 G.~Ecker, J.~Gasser, A.~Pich and E.~de Rafael,
  Nucl.\ Phys.\  B {\bf 321} (1989) 311.

\bibitem{GP06}
 M.~Golterman and S.~Peris,
  Phys.\ Rev.\ D {\bf 74}, 096002 (2006)
  [arXiv:hep-ph/0607152].


\bibitem{swiss2}
 G.~Ecker, J.~Gasser, H.~Leutwyler, A.~Pich and E.~de Rafael,
  Phys.\ Lett.\  B {\bf 223} (1989) 425;
for an update, see the last paper in Ref. \cite{MHA}. See also J.~Bijnens and E.~Pallante,
  Mod.\ Phys.\ Lett.\  A {\bf 11}, 1069 (1996)
  [arXiv:hep-ph/9510338].

\bibitem{Moussallam}
B.~Moussallam,
  Nucl.\ Phys.\  B {\bf 504} (1997) 381
  [arXiv:hep-ph/9701400].



\bibitem{MHA}
M.~Knecht and E.~de Rafael,
  Phys.\ Lett.\ B {\bf 424}, 335 (1998)
  [arXiv:hep-ph/9712457];
S.~Peris, M.~Perrottet and E.~de Rafael,
  JHEP {\bf 9805}, 011 (1998)
  [arXiv:hep-ph/9805442];
E.~de Rafael,
  Nucl.\ Phys.\ Proc.\ Suppl.\  {\bf 119} (2003) 71
  [arXiv:hep-ph/0210317];
  S.~Peris,
  arXiv:hep-ph/0204181;
 A.~Pich,
  Int.\ J.\ Mod.\ Phys.\ A {\bf 20} (2005) 1613
  [arXiv:hep-ph/0410322].




\bibitem{theworks}
S.~Peris, M.~Perrottet and E.~de Rafael,
  Phys.\ Lett.\  B {\bf 355} (1995) 523
  [arXiv:hep-ph/9505405];
 M.~Knecht, S.~Peris, M.~Perrottet and E.~de Rafael,
  Phys.\ Rev.\ Lett.\  {\bf 83}, 5230 (1999)
  [arXiv:hep-ph/9908283];
 M.~F.~L.~Golterman and S.~Peris,
  Phys.\ Rev.\  D {\bf 61} (2000) 034018
  [arXiv:hep-ph/9908252];
 M.~Knecht, S.~Peris and E.~de Rafael,
  Phys.\ Lett.\  B {\bf 508}, 117 (2001)
  [arXiv:hep-ph/0102017];
M.~Knecht and A.~Nyffeler,
  Eur.\ Phys.\ J.\  C {\bf 21} (2001) 659
  [arXiv:hep-ph/0106034];
M.~Knecht, S.~Peris, M.~Perrottet and E.~De Rafael,
  JHEP {\bf 0211} (2002) 003
  [arXiv:hep-ph/0205102];
M.~Golterman and S.~Peris,
  Phys.\ Rev.\  D {\bf 68} (2003) 094506
  [arXiv:hep-lat/0306028];
 T.~Hambye, S.~Peris and E.~de Rafael,
  JHEP {\bf 0305} (2003) 027
  [arXiv:hep-ph/0305104];
V.~Cirigliano, G.~Ecker, M.~Eidemuller, A.~Pich and J.~Portoles,
  Phys.\ Lett.\  B {\bf 596} (2004) 96
  [arXiv:hep-ph/0404004];
J.~Bijnens, E.~Gamiz and J.~Prades,
  Nucl.\ Phys.\ Proc.\ Suppl.\  {\bf 133} (2004) 245
  [arXiv:hep-ph/0309216];
V.~Cirigliano, G.~Ecker, M.~Eidemuller, R.~Kaiser, A.~Pich and J.~Portoles,
  JHEP {\bf 0504} (2005) 006
  [arXiv:hep-ph/0503108];
V.~Cirigliano, G.~Ecker, M.~Eidemuller, R.~Kaiser, A.~Pich and J.~Portoles,
  Nucl.\ Phys.\  B {\bf 753} (2006) 139
  [arXiv:hep-ph/0603205];
J.~Bijnens, E.~Gamiz and J.~Prades,
  JHEP {\bf 0603} (2006) 048
  [arXiv:hep-ph/0601197].









\bibitem{GP02}
M.~Golterman and S.~Peris,
Phys.\ Rev.\ D {\bf 67}, 096001 (2003) [arXiv:hep-ph/0207060].



\bibitem{Lipartia}
J.~Bijnens, E.~Gamiz, E.~Lipartia and J.~Prades,
  JHEP {\bf 0304}, 055 (2003)
  [arXiv:hep-ph/0304222].

\bibitem{Friot}
 S.~Friot, D.~Greynat and E.~de Rafael,
  JHEP {\bf 0410}, 043 (2004)
  [arXiv:hep-ph/0408281].

\bibitem{Baker}
G.A. Baker and P. Graves-Morris, \textit{Pade Approximants}, Encyclopedia of Mathematics and its
Applications, Cambridge Univ. Press 1996; G.A. Baker, \textit{Essentials of Pade Approximants},
Academic Press 1975.


\bibitem{meromorphic}
Third and fourth paper in \cite{MHA}.

\bibitem{Pommerenke} C. Pommerenke, \textit{Pade approximants and convergence in
capacity}, J. Math. Anal. Appl. \textbf{41} (1973) 775. Reviewed in Ref. \cite{Baker}, Section 6.5,
Theorem 6.5.4, Collorary 1, .


\bibitem{PerisPade}
S.~Peris,
  Phys.\ Rev.\ D {\bf 74} (2006) 054013
  [arXiv:hep-ph/0603190].

\bibitem{PerisPade2}
See the Table in Ref. \cite{PerisPade}.


\bibitem{Migdal}
 A.~A.~Migdal,
  Annals Phys.\  {\bf 109} (1977) 365.

\bibitem{oscar}
O.~Cata,
  arXiv:hep-ph/0701196;
 O.~Cata,
  arXiv:hep-ph/0605251.

\bibitem{spectra}
See, e.g.,  M.~Golterman and S.~Peris,
  JHEP {\bf 0101} (2001) 028
  [arXiv:hep-ph/0101098];
 A.~A.~Andrianov, S.~S.~Afonin, D.~Espriu and V.~A.~Andrianov,
  Int.\ J.\ Mod.\ Phys.\  A {\bf 21} (2006) 885
  [arXiv:hep-ph/0509144].





\bibitem{Baker3}
First book in Ref. \cite{Baker}; Section 5.5.



\bibitem{Erlich}
 J.~Erlich, G.~D.~Kribs and I.~Low,
  Phys.\ Rev.\  D {\bf 73} (2006) 096001
  [arXiv:hep-th/0602110] and references therein.


\bibitem{duality}
 O.~Cata, M.~Golterman and S.~Peris,
  JHEP {\bf 0508}, 076 (2005)
  [arXiv:hep-ph/0506004].

\bibitem{Donoghue}
Discussion session led by J. Donoghue at the workshop ``Matching light quarks to hadrons'',
Benasque Center for Science, Benasque, Spain, July-August 2004,
http://benasque.ecm.ub.es/2004quarks/2004quarks.htm



\bibitem{Baker2}
See the second book in \cite{Baker}, Chapter 14, Corollary 14.3.



\bibitem{PGV}C. Diaz-Mendoza, P. Gonzalez-Vera and R. Orive, Appl. Num. Math.
\textbf{53} (2005) 39 and references therein.



\bibitem{Shifman}
M.~A.~Shifman,
  arXiv:hep-ph/0009131.





\bibitem{FESR}
J.~Rojo and J.~I.~Latorre,
  JHEP {\bf 0401}, 055 (2004)
  [arXiv:hep-ph/0401047];
J.~Bijnens, E.~Gamiz and J.~Prades,
  JHEP {\bf 0110} (2001) 009
  [arXiv:hep-ph/0108240];
S.~Peris, B.~Phily and E.~de Rafael,
  Phys.\ Rev.\ Lett.\  {\bf 86} (2001) 14
  [arXiv:hep-ph/0007338];

\bibitem{pw}
V.~Cirigliano, E.~Golowich and K.~Maltman,
  Phys.\ Rev.\  D {\bf 68} (2003) 054013
  [arXiv:hep-ph/0305118];
C.~A.~Dominguez and K.~Schilcher,
  Phys.\ Lett.\  B {\bf 581} (2004) 193
  [arXiv:hep-ph/0309285].

\bibitem{Narison}
 S.~Narison,
  Phys.\ Lett.\  B {\bf 624}, 223 (2005)
  [arXiv:hep-ph/0412152];



\bibitem{almasy}
 A.~A.~Almasy, K.~Schilcher and H.~Spiesberger,
  arXiv:hep-ph/0612304.

\bibitem{Regge}
G.~'t Hooft,
Nucl.\ Phys.\ B {\bf 75} (1974) 461;
C.~G.~Callan, N.~Coote and D.~J.~Gross,
Phys.\ Rev.\ D {\bf 13}, 1649 (1976);
M.~B.~Einhorn,
Phys.\ Rev.\ D {\bf 14} (1976) 3451.




\bibitem{KPdeR}
 M.~Knecht, S.~Peris and E.~de Rafael,
  Phys.\ Lett.\ B {\bf 443}, 255 (1998)
  [arXiv:hep-ph/9809594].

\bibitem{BK} S.~Peris and E.~de Rafael,
  Phys.\ Lett.\  B {\bf 490} (2000) 213
  [arXiv:hep-ph/0006146];
  O.~Cata and S.~Peris,
  JHEP {\bf 0407}, 079 (2004)
  [arXiv:hep-ph/0406094];
  O.~Cata and S.~Peris,
  JHEP {\bf 0303}, 060 (2003)
  [arXiv:hep-ph/0303162].

\bibitem{subleading}
O.~Cata and S.~Peris,
  Phys.\ Rev.\  D {\bf 65} (2002) 056014
  [arXiv:hep-ph/0107062];
 I.~Rosell, P.~Ruiz-Femenia and J.~Portoles,
  JHEP {\bf 0512} (2005) 020
  [arXiv:hep-ph/0510041];
J.~Portoles, I.~Rosell and P.~Ruiz-Femenia,
  arXiv:hep-ph/0611375;
I.~Rosell, J.~J.~Sanz-Cillero and A.~Pich,
  JHEP {\bf 0408} (2004) 042
  [arXiv:hep-ph/0407240].

\bibitem{PDG}
 W.~M.~Yao {\it et al.}  [Particle Data Group],
  J.\ Phys.\ G {\bf 33} (2006) 1.

\bibitem{L10}
 M.~Davier, L.~Girlanda, A.~Hocker and J.~Stern,
  Phys.\ Rev.\  D {\bf 58}, 096014 (1998)
  [arXiv:hep-ph/9802447].

\bibitem{ecker}
 V.~Cirigliano, G.~Ecker, M.~Eidemuller, R.~Kaiser, A.~Pich and J.~Portoles,
  JHEP {\bf 0504} (2005) 006
  [arXiv:hep-ph/0503108].




\bibitem{LRoos}
H.~Leutwyler and M.~Roos,
  Z.\ Phys.\  C {\bf 25} (1984) 91;
 M.~Jamin, J.~A.~Oller and A.~Pich,
  JHEP {\bf 0402} (2004) 047
  [arXiv:hep-ph/0401080];
 D.~Becirevic {\it et al.},
  Nucl.\ Phys.\  B {\bf 705} (2005) 339
  [arXiv:hep-ph/0403217];
D.~J.~Antonio {\it et al.},
  arXiv:hep-lat/0702026.




\end{thebibliography}
\end{document}